# Dielectric behavior of Copper Tantalum Oxide

B. Renner[1], P. Lunkenheimer[2], M. Schetter[2], A. Loidl[2], A. Reller[1] and S. G. Ebbinghaus[1]

[1] Festkörperchemie, Institut für Physik, Universität Augsburg, D-86159 Augsburg, Germany

[2] Experimentalphysik V, Elektronische Korrelationen und Magnetismus, Institut für Physik, Universität Augsburg, D-86159 Augsburg, Germany

**Abstract**

A thorough investigation of the dielectric properties of $Cu_2Ta_4O_{12}$, a material crystallizing in a pseudo-cubic, perovskite-derived structure is presented. We measured the dielectric constant and conductivity of single crystals in an exceptionally broad frequency range up to GHz frequencies and at temperatures from 25 – 500 K. The detected dielectric constant is unusually high (reaching values up to $10^5$) and almost constant in a broad frequency and temperature range. $Cu_2Ta_4O_{12}$ possesses a crystal structure similar to $CaCu_3Ti_4O_{12}$, the compound for which such an unusually high dielectric constant was first observed. An analysis of the results using a simple equivalent circuit and measurements with different types of contact revealed that extrinsic interfacial polarization effects, derived from surface barrier capacitors are the origin of the observed giant dielectric constants. The intrinsic properties of $Cu_2Ta_4O_{12}$ are characterized by a (still relatively high) dielectric constant in the order of 100 and by charge transport via hopping conduction of Anderson-localized charge carriers.

**Introduction**

Materials with high dielectric constants are widely used in technological applications, like wireless communication systems such as cellular phones and global positioning systems, namely as capacitors, resonators or filters. High dielectric constants allow smaller capacitive components, thus offering the opportunity to decrease the size of electronic devices[1]. The demand to downscale integrated circuit technology requires the replacement of the currently used $SiO_2$ gate dielectrics by an alternative high-$\varepsilon'$ material. Recently, the discovery of giant dielectric constants in $CaCu_3Ti_4O_{12}$ was reported[2]. The $\varepsilon'$-values of this oxide exhibit only small variations with temperature and frequency, both being important constraints for technological applications. The explanation of this so far unique effect is still disputed. Ramirez *et al.*[3] excluded ferroelectricity as a cause of the unusual high dielectric response. Instead they explained their results by a relaxor model. A fit of the real part of the dielectric response as a function of frequency in this model leads to a high-frequency value of $\varepsilon_\infty = 80$.[4] Sinclair et al.[5] and Lunkenheimer *et al.*[6], on the other hand, demonstrated how apparent giant dielectric constants ($\varepsilon' > 1,000$) can be explained by simple capacitive effects resulting from depletion layers at grain boundaries or at sample contacts. Such devices, which are termed surface barrier layer capacitor (SBLC) or internal barrier layer capacitor (IBLC) lead to high effective permittivity values. As the highest dielectric constants in $CaCu_3Ti_4O_{12}$ were found in single crystal samples, only IBLCs resulting from twin boundaries[2] or SBLCs have to be considered. However, in a recent publication an intrinsic origin of the detected giant dielectric constant was suggested, and a theoretical explanation in terms of relaxing defects was proposed by Ramirez and coworkers[7]. This interpretation was substantiated by the fact that the isostructural compound $CdCu_3Ti_4O_{12}$ reveals a much lower dielectric constant[8]. However the number of puzzling questions was even increased by the fact that from first principle calculations the lattice dynamics of $CaCu_3Ti_4O_{12}$ and $CdCu_3Ti_4O_{12}$ look very similar[9].

Until now, $CaCu_3Ti_4O_{12}$ was the only member of the class of $ACu_3M_4O_{12}$-oxides (A = alkali, alkaline earth, rare earth metal or vacancy, M = transition metal) to exhibit such an unusual high dielectric constant[10], known so far only from ferroelectrics or IBLCs. In the present report we demonstrate that $Cu_2Ta_4O_{12}$ exhibits a dielectric behavior similar to the one of $CaCu_3Ti_4O_{12}$. The crystal structure of $Cu_2Ta_4O_{12}$ can be derived from the $CaCu_3Ti_4O_{12}$ structure by leaving the Ca sites unoccupied and replacing 1/3 of the copper sites by vacancies. Compared to the ideal cubic perovskite structure, the $MO_6$-octahedra in the copper tantalate are rotated along the spatial diagonals (<111> direction) leading to a rectangular four-fold oxygen co-ordination of the copper ions (Figure 1). We present a thorough dielectric characterization of this material in a broad temperature and frequency range up to 3 GHz and clarify the question of the intrinsic nature of the observed dielectric response.

**Experimental Section**

Single crystals of $Cu_2Ta_4O_{12}$ were grown from a self-flux of copper oxide according to the procedure described elsewhere [11]. The obtained crystals were black rectangular plates up to several mm in diameter. These crystals possess a pseudo-cubic unit cell with $a \approx 7.5$ Å. Synchrotron X-ray diffraction measurements on crushed crystals, on the other hand, revealed deviations from cubic symmetry in the order of 1‰, which is in very good agreement with previous reports[11].

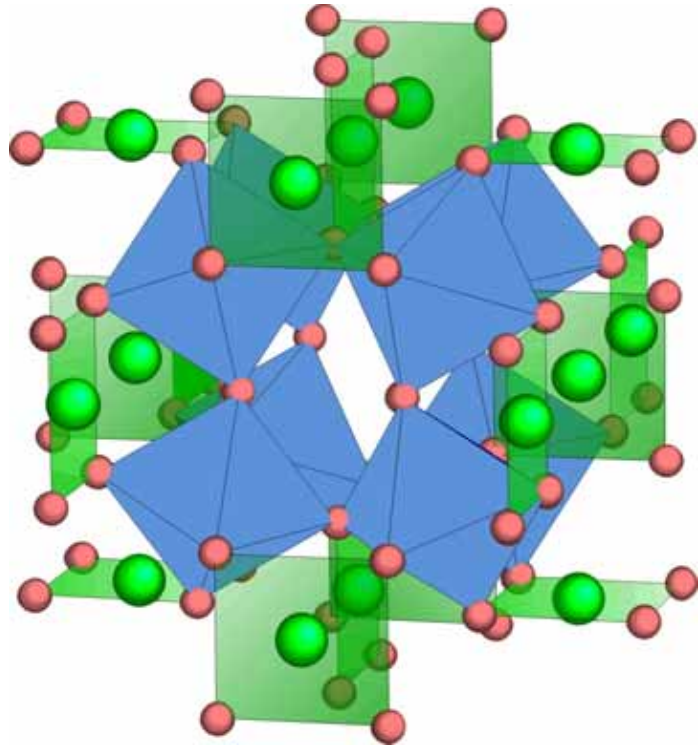

**Figure 1.** Crystal structure of $Cu_2Ta_4O_{12}$. $TaO_6$-octahedra (blue), Cu atoms (green), and oxygen atoms (red) of $Cu_2Ta_4O_{12}$ are illustrated. 1/3 of the copper sites remain unoccupied.

The conductivity and permittivity of two single crystals were measured in the frequency range from 20 Hz up to 3 GHz. Crystal 1 was investigated at temperatures from 125 K to 500 K and crystal 2 at 20 K to 300 K. For the measurements an autobalance bridge (HP 4284A) was used at $\nu < 1$ MHz and a reflectometric technique, employing an impedance analyzer (HP 4291B and Agilent E 4991A), at $\nu > 1$ MHz[12]. From the measured capacitance and conductance, the dielectric constant and conductivity were calculated using the geometry of the parallel-plate capacitor formed by applying silver paint to adjacent sides of the single crystals. To check for the possible formation of SBLCs, for crystal 2 additional measurements were performed after removing the silver-paint contacts in an ultrasonic bath and replacing them by sputtered silver contacts of 100 nm thickness.

**Results and Discussion**

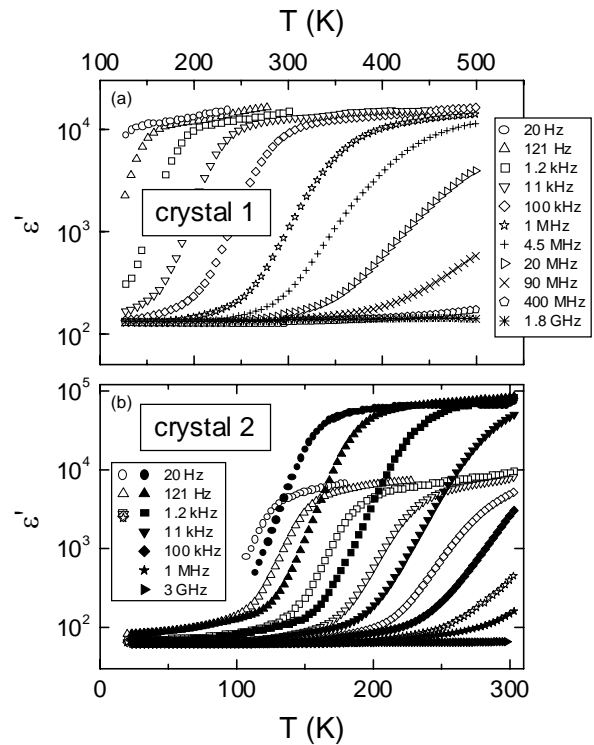

**Figure 2.** Temperature-dependence of the dielectric constant of two crystals of $Cu_2Ta_4O_{12}$ at different frequencies. The measurements reveal a step like transition from a value of $\varepsilon'_{low} \approx 10^2$ to a giant value of the dielectric constant of $\varepsilon'_{high} \approx 10^4$-$10^5$ at high temperatures. The step shifts to higher temperatures with increasing frequency. The measurements of crystal 1 (a) were obtained with silver-paint contacts. Crystal 2 (b) was contacted both with silver paint (open symbols) and with sputtered silver (closed symbols), revealing the extrinsic origin of the huge dielectric constants.

Figure 2 shows the temperature dependence of the dielectric constant for frequencies up to 3 GHz. We analyzed two different crystals: Figure 2a shows the results for crystal 1 measured at higher temperatures (125 K – 500 K) with silver-paint contacts while Figure 2b exhibits the data obtained for crystal 2 below room temperature with two types of contacts. Giant dielectric constants of up to $\varepsilon'_{high} \approx 10^4$ for the silver-paint (open symbols) contacts and $\varepsilon'_{high} \approx 10^5$ for the sputtered contacts (closed symbols) are detected at high temperatures and/or low frequencies. With decreasing temperature, step-like transitions to a lower value of $\varepsilon'_{low} \approx 10^2$ are observed, the transition temperatures shifting to lower values with decreasing frequency. The overall behavior closely resembles that reported for CaCu$_3$Ti$_4$O$_{12}$[2-5,7], with stated values of $\varepsilon'_{high}$ varying between 5.000 and 70.000 and $\varepsilon'_{low} \approx 100$ (±20). At frequencies higher than approximately 1 MHz, $\varepsilon'_{high}$ is not reached even for the measurements up to $T = 500$ K performed for crystal 1 (Figure 2a). Here, for $\nu = 1.8$ GHz the value of $\varepsilon' \approx 130$ was found to be almost temperature independent, retaining the value of $\varepsilon'_{low}$ for the whole temperature range investigated. Indications for a similar reduction of $\varepsilon'$ at high frequencies were also found in earlier results for CaCu$_3$Ti$_4$O$_{12}$, restricted to frequencies below 1 MHz[5], which is corroborated by the recent results at frequencies up to GHz from our group[13]. Irrespective of a possible non-intrinsic origin of the detected high $\varepsilon'$ values, this reduction certainly limits the applicability of these giant dielectric constant materials for high-frequency devices. A value of $\varepsilon'$ in the order of 100 is, on the other hand, still quite high compared to other technically relevant materials like $Ta_2O_5$ ($\varepsilon' \approx 26$) and $HfO_2$ ($\varepsilon' \approx 25$), both being prime candidates to replace $SiO_2$ as gate dielectrics in the near future. Additionally, the temperature dependence of the dielectric constant in the GHz regime is extremely small. This is an important property especially in frequency filters, making $Cu_2Ta_4O_{12}$ a promising material for such devices. Figure 2b clearly demonstrates a strong impact of the type of sample contact on the value of $\varepsilon'_{high}$: With sputtered silver much higher values are detected than for contacts formed by silver paint. In contrast, the value of $\varepsilon'_{low}$ is not affected by the type of contact. This result clearly proves the non-intrinsic nature of the measured giant values of the dielectric constant, giving strong hints for an SBLC mechanism as will be discussed in detail below.

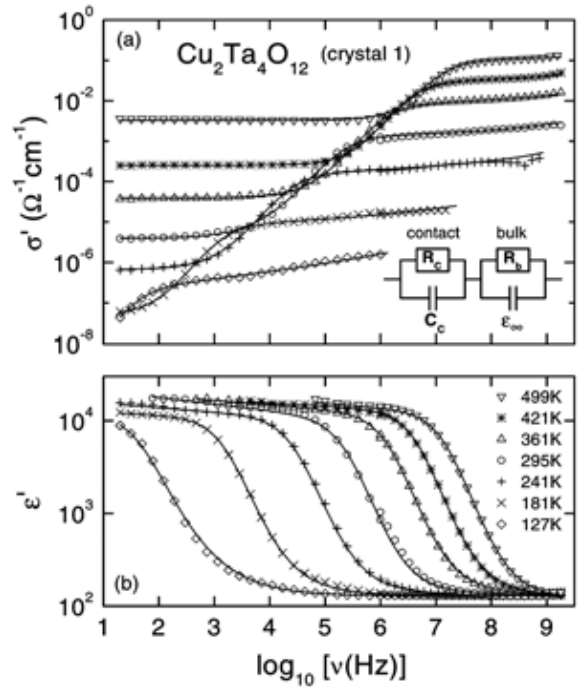

**Figure 3.** Frequency-dependent conductivity (a) and dielectric constant (b) at different temperatures of $Cu_2Ta_4O_{12}$ (crystal 1, silver-paint contacts). The solid lines are fits using the equivalent circuit

indicated in the figure, taking into account contributions from interfacial polarization.

Figures 3 and 4 show the real part of the conductivity $\sigma'$ (a) and the dielectric constant $\varepsilon'$ (b) as a function of frequency for crystal 1 with silver-paint contacts and for crystal 2 with sputtered silver contacts, respectively. The frequency dependence of the dielectric constant closely resembles the one observed in $CaCu_3Ti_4O_{12}$[4], showing sigmoidal curves that shift through the frequency window with temperature and revealing giant values for $\varepsilon'$ at low frequencies. In these plots, the quantities $\varepsilon'_{high}$ and $\varepsilon'_{low}$ show up as limiting values at low and high frequencies, respectively. For a possible explanation of the observed behavior, Maxwell-Wagner-type interfacial relaxation processes were considered[5,6]. Following this notion, the spectra were fitted with the simple equivalent circuit presented as insets in the figures. This circuit consists of a leaky capacitor connected in series with the bulk sample[14]. The former represents the most common way to model contributions from interfacial polarization processes. In case of SBLCs it can arise from the formation of Schottky barriers (depletion layers) in the semiconducting material close to the metallic electrodes. Additionally, an accumulation of defects or deviations from stoichiometry near the sample surface may contribute to such a capacitive surface layer. These thin layers of small conductivity act as a high capacitance in parallel with a large resistor, connected in series to the bulk sample. In case of IBLCs, the same type of contribution could emerge from grain boundaries in polycrystalline samples and possibly also from twin boundaries in single crystals. The bulk response can also be approximated by a leaky capacitor, with a bulk capacitance determined by the intrinsic dielectric constant $\varepsilon_\infty$ and a bulk resistance, which, depending on the charge transport process, may be complex and frequency-dependent.

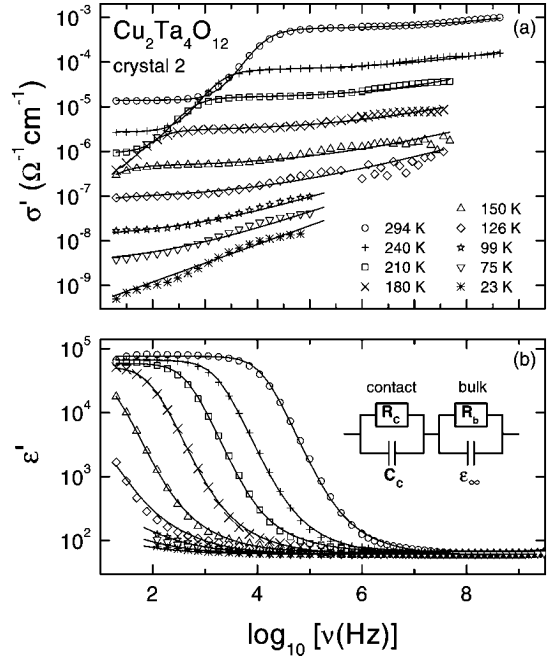

**Figure 4.** Frequency-dependent conductivity (a) and dielectric constant (b) at different temperatures of $Cu_2Ta_4O_{12}$ (crystal 2, sputtered silver contacts). The solid lines are fits using the equivalent circuit indicated in the figure, taking into account contributions from interfacial polarization.

The fits using this equivalent circuit – performed simultaneously for conductivity and dielectric constant – lead to a good description of the experimental data as demonstrated by the lines in Figures 3 and 4. To take into account the somewhat broadened steps in $\varepsilon'(\nu)$, in some cases a Cole-Cole distribution of the time constant of the leaky capacitor representing the barrier had to be used[14]. This can be ascribed to the roughness of the sample surface for the SBLC and to a distribution of twin boundaries for the IBLC case. Within this framework, the occurrence of apparently giant dielectric constants in the order of $10^4$, respectively $10^5$ at frequencies below 1 MHz reflects the fact that the low-frequency response of the circuit is dominated by the very thin interface capacitors, reaching high values of up to several nF. By increasing the frequency, the interfacial capacitance begins to act like a short and the intrinsic bulk-response is detected. It is characterized by a much smaller capacitance of about 3-10 pF, corresponding to an intrinsic $\varepsilon_\infty \approx 10^2$. The step in the dielectric response is accompanied by an increase in conductivity of about two orders of magnitude due to the successive shorting of the high interface resistance and the approach of the intrinsic bulk conductivity at high frequencies. Via the time constant of the circuit, the semiconducting characteristics of the bulk conductivity leads to the observed strong temperature shift of the step.

The intrinsic bulk conductivity, which can be read off at high frequencies, exhibits a slight increase with frequency, becoming steeper for low temperatures (Figure 4a). It is taken into account in the fits by a power law with exponent smaller than one (and the corresponding contribution in $\varepsilon'(\nu)$ (ref. 14), in addition to a frequency-independent dc contribution, $\sigma_{dc}$. This leads to an overall bulk conductivity of the form $\sigma_{dc} + \sigma_0 \nu^s$ (s<1), a behavior that is observed in a variety of materials including doped semiconductors and numerous transition metal oxides (see, e.g. refs. 14-18). For conducting materials, the $\nu^s$ power law is a hallmark of hopping of Anderson-localized charge carriers. There is a variety of theoretical approaches that deduce this behavior from microscopic transport mechanisms, involving hopping over or tunneling through an energy barrier, separating different localized sites[15]. A prerequisite for Anderson-localization is the presence of disorder, as prevailing, e.g., in amorphous or doped semiconductors. For the case of $Cu_2Ta_4O_{12}$, it is obvious that a considerable amount of disorder is generated by the statistical distribution of the vacancies on the copper sites. The fits of the data shown in Figures 3 and 4 reveal an exponent $s$ varying between 0.25 and 0.5, which is relatively small compared to other materials where in most cases values above 0.5 are found.[14] However, except for the lowest temperatures (Figure 4a), the bulk response seems to be determined to a large extent by the dc component and to obtain precise information on the absolute value of $s$, measurements up to higher frequencies would be necessary.

In Figure 4a, in the bulk-dominated region and especially at low temperatures small deviations between experimental data and fits are noted. Namely, the data show the slight indication of a second step-like feature (its point of inflection being located e.g. at $10^{3.5}$ Hz for 75 K), which is not covered by the fits. This feature, becoming most evident at the lowest temperatures where the experimental device operates at its resolution limit, may be an artifact but it may also indicate a further relaxational contribution, either intrinsic, or due to internal barriers, e.g. twin boundaries.

In Figure 5, a direct comparison of the results of crystal 2 performed with silver-paint contacts (open symbols) and sputtered silver contacts (filled symbols) is shown. The figure strongly corroborates the picture developed above, assuming a non-intrinsic origin of the giant $\varepsilon'$-values: For high frequencies both contact types lead to identical values of conductivity and dielectric constant because their contributions no longer play a role due to the shortening of the contact capacitor. At low frequencies the different contact types cause a strong deviation in $\varepsilon'$, which clearly proves the formation of SBLCs leading to the detected giant dielectric constants. The fact that the sputtered contacts lead to much higher values of $\varepsilon'$, strongly points to the SBLCs being formed by Schottky barriers at the contact-sample interface. Namely, it seems reasonable that sputtered contacts give rise to a more effective formation of the Schottky barriers because the very small metal clusters applied during sputtering will lead to a larger area of direct metal-semiconductor contact than for the relatively large particles ($\geq\mu$m) suspended in the silver paint.

The question arises why $Cu_2Ta_4O_{12}$ is only the second material in the series of $ACu_3M_4O_{12}$-oxides to exhibit such a high apparent dielectric constant. In reference[10], dielectric constants measured at 100 kHz and room temperature were reported for 21 further members of this group of materials, revealing a broad range of values between $\varepsilon' \approx 33$ and $\varepsilon' \approx 3,560$. Within the SBLC framework, a rather trivial explanation of this finding would be different surface treatments, types of contacts, etc., which certainly affect the formation of the SBLCs, and thus have a strong impact on the detected values. One may also suspect that material properties, e.g. different electron work functions, leading to a different thickness of the Schottky-diode depletion layers, might play an important role here. In addition, variations in the intrinsic bulk resistivities may lead to a shift of the $\varepsilon'(\nu)$-steps to lower or higher frequencies.

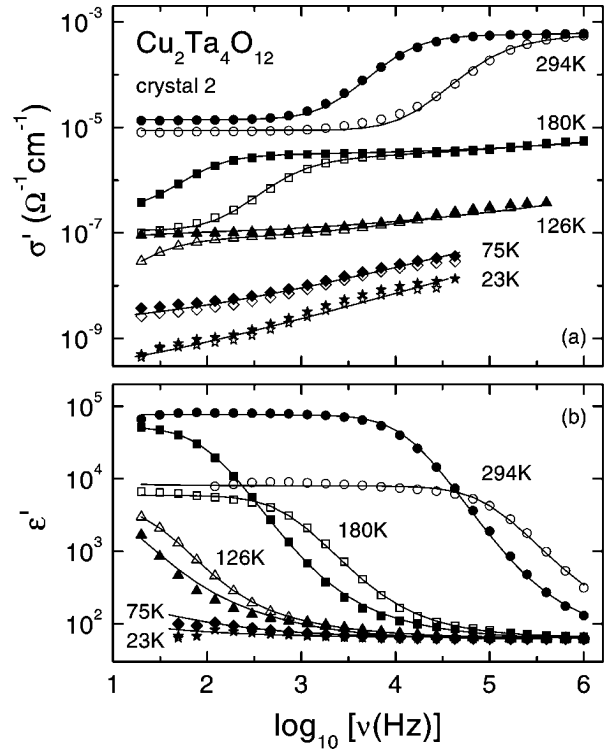

**Figure 5.** Frequency-dependent conductivity (a) and dielectric constant (b) at different temperatures of $Cu_2Ta_4O_{12}$ (crystal 2) with contacts formed by silver paint (open symbols) and sputtered silver (closed symbols). The solid lines are fits using the equivalent circuit indicated in Fig. 4. The large difference in the low-frequency dielectric constant points to SBLCs causing the giant values of $\varepsilon'$.

The fits of the spectra using the equivalent circuit, shown in Figs. 3 and 4 allow us to determine the intrinsic dc-conductivity, $\sigma_{dc}$. Figure 6 shows the resulting temperature dependence of $\sigma_{dc}$ for both crystals. While there is a small difference between the results for crystal 1 and 2, which may be caused by a slightly different stoichiometry, the overall behavior is very similar. Namely, while $\sigma_{dc}(T)$ exhibits semiconducting characteristics (i.e. $\sigma_{dc}$ increases with temperature), it strongly deviates from simple thermally activated behavior ($\sigma_{dc} \sim \exp[-E/(k_B T)]$ with $E$ the gap energy). This can be deduced from the clear deviations from straight-line behavior in the Arrhenius representation, given in the inset of Fig. 6 (squares, lower scale). Only at temperatures above about 200 K ($1000/T < 5$) Arrhenius behavior is fulfilled, revealing an energy barrier of about 0.24eV (solid line). In the inset of Fig. 6 the same data are plotted also in a representation that should linearize the data for a behavior $\sigma_{dc} \sim \exp[-(T_0/T)]^{1/4}$ (triangles, upper scale). Such a behavior is predicted for 3D variable range hopping (VRH), involving the phonon-assisted tunneling of Anderson-localized charge carriers[19]. Obviously, the VRH prediction provides a reasonable description of $\sigma_{dc}(T)$ in $Cu_2Ta_4O_{12}$ over a broad temperature range (dashed line). We obtain values of $T_0 \approx 6 \times 10^8$ K, which is of similar magnitude as observed in various other transition metal oxides[17,18,20].

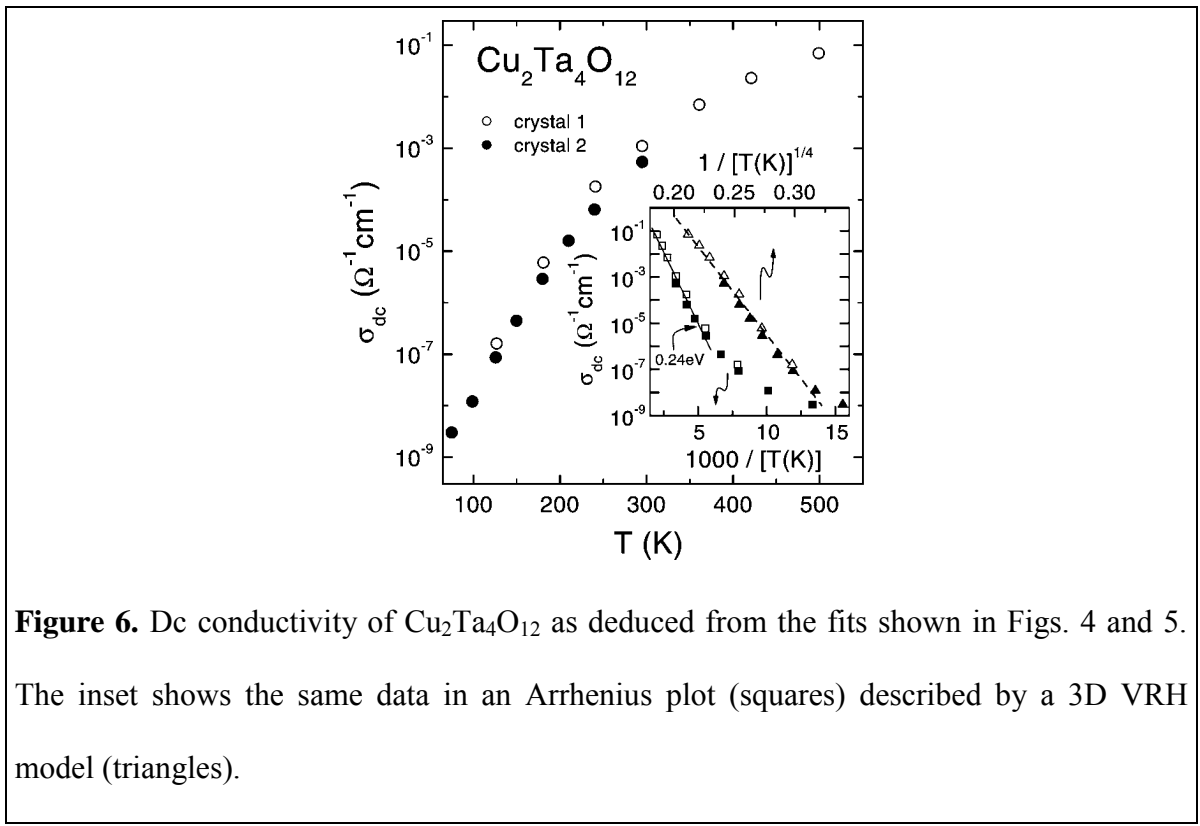

**Figure 6.** Dc conductivity of $Cu_2Ta_4O_{12}$ as deduced from the fits shown in Figs. 4 and 5. The inset shows the same data in an Arrhenius plot (squares) described by a 3D VRH model (triangles).

**Conclusions**

In the present work we showed that at least one other member of the $ACu_3M_4O_{12}$-oxides also exhibits a giant dielectric constant. An intrinsic mechanism causing the unusual dielectric behavior of these two materials is not nessessary. Our measurements using two different ways of contact preparation strongly suggest that the formation of SBLCs at the sample-electrode interface is the more likely origin of the detected giant dielectric constants.

Performing a careful analysis of the experimental spectra via an equivalent circuit analysis reveals that the intrinsic charge transport properties in $Cu_2Ta_4O_{12}$ are dominated by hopping conduction of Anderson-localized charge carriers. Evidence for Anderson localization is consistently revealed by both, the ac- and the dc-conductivity. It seems reasonable that the localization of the charge carriers is induced by the statistical distribution of the vacancies on the copper sites of $Cu_2Ta_4O_{12}$. Our results reveal an intrinsic dielectric constant of $Cu_2Ta_4O_{12}$ of about 100. This value is relatively high and shows that it is worthwhile to further investigate the dielectric properties of the $ACu_3M_4O_{12}$-oxides. An optimization of composition might lead to sufficiently high values to establish these materials as a new generation of dielectrics for capacitive circuit elements in electronic applications.

**Acknowledgements**

This research was supported by the Deutsche Forschungsgemeinschaft via the Sonderforschungsbereich 484 and partly by the BMBF via VDI/EKM, FKZ 13N6917.

**References and Notes**

1. Cava, R. J. *J. Mater. Chem.* **2001**, *11*, 54-62.

2. Subramanian, M. A.; Li, D.; Duan, N.; Reisner, B. A.; Sleight, A. W. *J. Solid State Chem.* **2000**, *151*, 323-325.

3. Ramirez A. P; Subramanian, M. A.; Gardel, M.; Blumberg, G.; Li, D.; Vogt, T.; Shapiro, S. M *Solid State Commun.* **2000**, *115*, 217-220.

4. Homes, C. C.; Vogt, T.; Shapiro, S . M.; Wakimoto, S.; Ramirez, A. P. *Science* **2001**, *293*, 673-676.

5. Sinclair, D. C.; Adams, T. B.; Morrison, F. D.; West, A. R. *Appl. Phys. Lett.* **2002**, *80* (12), 2153-2155.

6. Lunkenheimer P.; Bobnar, V.; Pronin, A. V.; Ritus, A. I.; Volkov, A. A.; Loidl A. *Phys. Rev. B* **2002**, *66*, 052105.

7. Ramirez, A. P.; Lawes, G.; Butko, V.; Subramanian, M. A.; Varma, C. M. *cond-mat/0209498* (2002).

8. Homes, C. C; Vogt, T.; Shapiro, S. M.; Wakimoto, S.; Subramanian, M. A.; Ramirez, A. P. *Phys. Rev. B* **2003**, *67*, 092106.

9. He, L.; Neaton, J. B.; Vanderbilt, D.; Cohen, M. H. *Phys. Rev. B* **2003**, *67*, 012103.

10. Subramanian, M. A.; Sleight, A. W. *Solid State Sci.* **2002**, *4*, 347-351.

11. Vincent, H.; Bochu, B.; Aubert, J. J.; Joubert, J. C.; Marezio, M. *J. Solid State Chem*. **1978**, *24*, 245-253.

12. Böhmer, R.; Maglione, M.; Lunkenheimer, P.; Loidl, A. *J. Appl. Phys.* **1989**, 65, 901-904; Schneider, U.; Lunkenheimer, P.; Pimenov, A.; Brand, R.; Loidl, A. Ferroelectrics **2001**, 249, 89-98

13. P. Lunkenheimer *et al.*, unpublished results.

14. Jonscher, A.K. *Dielectric Relaxations in Solids*; Chelsea Dielectrics Press: London, 1983; Jonscher, A. K. *Universal Relaxation Law*; Chelsea Dielectrics Press: London, 1996.

15. Elliott, S. R. **1987**, 36, 135-218 (1987); Long, A. R. *Adv. Phys.* **1982, 31,** 553-637 (1982).

16. Lunkenheimer, P.; Resch, M.; Loidl, A.; Hidaka, Y. *Phys. Rev. Lett.* **1992**, 69, 498-501; Bobnar, V.; Lunkenheimer, P.; Paraskevopoulos, M.; Loidl, A. *Phys. Rev. B* **2002**, 65, 184403.

17. Seeger, A.; Lunkenheimer, P.; Hemberger, J.; Mukhin, A. A.; Ivanov, V. Yu.; Balbashov, A. M.; Loidl, A. *J. Phys.: Condens. Matter* **1999**, 11, 3273-3290.

18. Sichelschmidt, J.; Paraskevopoulos, M.; Brando, M.; Wehn, R.; Ivannikov, D.; Mayr, F.; Pucher, K.; Hemberger, J.; Pimenov, A.; Krug von Nidda, H.-A.; Lunkenheimer, P.; Ivanov, V. Yu.; Mukhin, A. A.; Balbashov, A. M.; Loidl, A. *Euro. Phys. J. B* **2001**, 20, 7-17.

19. N. F. Mott, N. F.; Davis, E. A. *Electronic Processes in Non-Crystalline Materials*; Clarendon Press: Oxford, 1979.

20. Fontcuberta, J.; Martinez, B.; Seffar, A.; Pinol, S.; Garcia-Muñoz, J.L.; Obradors, X. *Phys. Rev. Lett.* **1996,** 76, 1122-1125.